\documentclass[onecolumn, preprintnumbers, showpacs, aps]{revtex4}
\usepackage{graphicx}

\newcommand{\bi}{\bibitem}
\newcommand{\be}{\begin{eqnarray}}
\newcommand{\ee}{\end{eqnarray}}

\begin{document}

\title{Black holes as antimatter factories}

\author{Cosimo Bambi$^{\rm 1}$}
\author{Alexander D.~Dolgov$^{\rm 2,3,4}$}
\author{Alexey A.~Petrov$^{\rm 1,5}$}

\affiliation{$^{\rm 1}$Department of Physics and Astronomy, 
Wayne State University, Detroit, MI 48201, USA\\
$^{\rm 2}$Istituto Nazionale di Fisica Nucleare, 
Sezione di Ferrara, I-44100 Ferrara, Italy \\
$^{\rm 3}$Dipartimento di Fisica, 
Universit\`a degli Studi di Ferrara, I-44100 Ferrara, Italy \\
$^{\rm 4}$Institute of Theoretical and Experimental Physics,
113259 Moscow, Russia \\
$^{\rm 5}$Michigan Center for Theoretical Physics, 
University of Michigan, Ann Arbor, MI 48109, USA}

\date{\today}

\preprint{WSU-HEP-0808}

\begin{abstract}
We consider accretion of matter onto a low mass black hole 
surrounded by ionized medium. We show that, because of the
higher mobility of protons than electrons, the black hole would 
acquire positive electric charge. If the black hole's mass is 
about or below $10^{20}$~g, the electric field 
at the horizon can reach the critical value which leads to 
vacuum instability and electron--positron pair production by 
the Schwinger mechanism. Since the positrons are ejected by 
the emergent electric field, while electrons are back--captured, 
the black hole operates as an antimatter factory which 
effectively converts protons into positrons.
\end{abstract}


\maketitle


{\sc Introduction --} 
It is well-known fact that the difference between the mass of 
proton and electron can lead to predominant capture of protons by
celestial bodies, which can make them electrically charged~\cite{papers}. 
This process can be exemplified by the case of a cold object,
such as a planet, surrounded by an atmosphere of positive ions and 
electrons in thermal equilibrium. Here, the escape velocity can 
be reached easier by electrons, which thus would leave the atmosphere. 
This process leads to the appearance of growing electrostatic field around 
the planet, making electron escape more and more difficult. 
The equilibrium is eventually reached when the sums of all forces 
acting on ions and electrons are equal, which results in non-zero 
charge acquired by the planet. A similar mechanism can apply to a
radiating object. While protons and electrons are gravitationally 
attracted towards the radiating body with the same acceleration, 
the outgoing radiation acts differently on the two particle species.
Indeed, as the interaction between the outgoing electromagnetic 
radiation and charged particles is described by the Thompson
cross-section, the $e\gamma$ interaction dominates the $p\gamma$
one by a huge factor of $(m_p/m_e)^2 \sim 3\cdot 10^{6}$,
making proton-photon interaction effectively negligible.
The stationary solution is again the one in which the astrophysical 
body is electrically charged, so the emergent electrostatic field 
makes protons and electrons fall with the same acceleration. In the
standard case of radiative atmospheres the effect of the
accumulated charge is negligible, so the electric field is
not taken into account.

In ref.~\cite{turolla}, the authors discussed the case of a 
black hole (BH) which accreted spherically with luminosity 
close to the Eddington limit. If the BH mass is smaller than 
about $10^{20}$~g, the electrostatic field of the stationary 
configuration at horizon exceeds the critical limit to produce 
electron--positron pairs. The crucial ingredient of the mechanism 
is that BH accretion has to be close to the Eddington limit, 
because the outgoing radiation has to inhibit the electron accretion. 
This is only possible if the density of the surrounding matter 
is huge, roughly above $10^{24}$~protons/cm$^3 \approx 1$~g/cm$^3$. 
In this paper we revisit this mechanism, taking the particle
thermal distribution into account. We find that the process 
of proton--to--positron transformation can work even if the 
particle number density of the medium around the BH is much
lower, close to realistic values. Interesting phenomenological 
implications are therefore possible.


{\sc Schwinger mechanism at Schwarzschild horizon --}
Let us consider a BH of mass $M$ surrounded by plasma
of protons and electrons. In the simplest case of perfect 
spherical symmetry, the radial part of the equations of 
motion for the proton and electron fluids are respectively
\be
\label{dot-vp}
\dot v_p &=& -\frac{G_N M}{r^2} 
+ \frac{\alpha Q}{r^2 m_p} 
+ \frac{L \, \sigma_{\gamma p}}{4 \pi r^2 m_p} 
- \frac{\sigma_{\gamma p} n_\gamma \omega_\gamma}{m_p} \, v_p
- \frac{n_p \sigma_{pe} P}{m_p} \, (v_p - v_e) \, , \\
\dot v_e &=& -\frac{G_N M}{r^2} 
- \frac{\alpha Q}{r^2 m_e} 
+ \frac{L \, \sigma_{\gamma e}}{4\pi r^2 m_e} 
- \frac{\sigma_{\gamma e} n_\gamma \omega_\gamma}{m_e} \, v_e
+ \frac{n_e \sigma_{pe} P}{m_e} \, (v_p-v_e) \, .
\label{dot-ve} 
\ee
Here $v_p$ and $v_e$ are the proton and electron fluid velocities
(that is the average velocities of $p$ and $e$ in the plasma 
which do not include the large chaotic thermal velocities), $Q$ 
is the electric charge of the BH in proton charge units, 
$\alpha = e^2/4\pi = 1/137$, $\sigma_{ij}$ is the cross section of 
scattering of $i$ on $j$, $L$ is the luminosity in the comoving 
frame of the accretion flow, $P$ is the momentum transfer in 
$ep$--scattering, $n_p$ and $n_e$ are the number densities of 
$p$ and $e$ around the BH, $n_\gamma$ is the photon number 
density in the photon bath surrounding the BH and $\omega_\gamma$ 
is the photon energy, which is roughly the momentum transfer 
in $p\gamma$-- and $e\gamma$--scattering. We neglect the angular
momentum term because, as explained below, we are interested in
the particles with low angular momentum. 
Eqs.~(\ref{dot-vp}) and (\ref{dot-ve}) indeed recover the Eddington 
limit for $Q=0$ and stationary flux $\dot{v}_j=0$ if the fourth term 
on the right hand side of the two equations, that is, the term due to 
scattering on the thermal bath of photons, is neglected. 
Taking $Q=0$, $n_p = n_e$ and $\dot{v}_j=0$, we can solve the 
two equations for $L$, finding the usual Eddington luminosity, 
$L_E = 4 \pi G_N M m_p / \sigma_{e\gamma}$.

In what follows, it turns out to be useful to define the 
dimensionless quantities
\be
K = \frac{m_p}{m_e} \gg 1 \, , \qquad
q = \frac{\alpha Q}{r_g m_p} \qquad {\rm and} \qquad 
l = \frac{L \sigma_{p\gamma}}{4 \pi r_g m_p} \ll 1 \, ,
\ee
where $r_g = 2 G_N M$ is the gravitational radius of the BH.
Since the cross sections of $e\gamma$-- and $p\gamma$--scattering
is inversely proportional respectively to $m_e^2$ and $m_p^2$,
the impact of the radiation pressure on acceleration for protons
is suppressed with respect to electrons by $K^3$.

Let us consider particles inside the mean free path from 
the BH, $r < \lambda_j$ ($j=p,e$). If their velocity is 
small enough, i.e. $v_j < v_j^c (r) = \sqrt{F_j/r}$,
where $F_p = r_g(1 - 2q - 2l)$ and $F_e = r_g(1 + 2Kq - 2lK^3)$,
such particles would be captured by the BH. Even if we added the
angular momentum term into eqs.~(\ref{dot-vp}) and (\ref{dot-ve}),
which is about $J^2/r^3$ and has positive sign preventing 
accretion, the rather small velocities 
lead to a centrifugal to gravitational force ratio
\be
\frac{J^2/r^3}{G_N M/r^2} = \frac{r v^2_p}{G_N M} <
\frac{F_p}{G_N M} = 2(1-2q-2l)
\ee
for protons and $2(1+2Kq-2lK^3)$ for electrons. Since this ratio
for protons is always smaller than 2, the account of their angular
momentum cannot significantly change the protons' accretion, while
the electrons' accretion may noticeably slow down. This would increase
the charging rate. Then, most of the captured particles are quickly 
swallowed by the BH. Here indeed the picture is quite different
from the widely studied astrophysical accretion for large objects:
charged particles can easily lose angular momentum by emitting 
classical electromagnetic radiation due to accelerating motion 
around the BH and probably even by Coulomb scattering, because 
of the higher particle number density in the vicinity of the BH.

The particles propagate freely inside distance $r$, so we can 
assume thermal distribution and consider their motion under the 
influence of gravitational and Coulomb forces only. The number 
density of particles with velocity $v_j < v_j^c$ is 
\be
n^c_j (r) = C_j(r) \int^{v_j^c(r)}_0 dv v^2 
\exp\left(-\frac{m_j v^2}{2T}\right) \, .
\label{nc-of-r}
\ee
If the plasma were in equilibrium, $C_j(r)$ would be given by
$C_j(r) = C_j \exp\left(G_N M m_j/rT\right)$,
where $C_j$ is the value of $C_j(r)$ far from the BH. However,
in what follows we neglect the dependence of $C_j(r)$ on $r$ and we 
take $C_j(r) = C_j$: the BH accretion is actually determined
by the particle distribution far from the BH, where the chemical
potential is small, while in the neigborhood of the BH, protons
and electrons are not in equilibrium, since there is
an intense ingoing flux to the BH. The quantity $C_j$ can be 
determined by the normalization condition
\be
n_j = C_j \int_0^\infty dv v^2 
\exp\left(-\frac{m v^2}{2T}\right)
= C_j \frac{\sqrt{\pi}}{4}
\left(\frac{2 T}{m}\right)^{3/2}  \, , 
\label{n}
\ee
where $n_j$ is the particle number density at infinity and we
can reasonably assume that $n_p = n_e$.

The total number of particles with $v_j < v_j^c$ inside the shell 
from $r$ to $r+dr$ is
%
$dn_j(r) = 4 \pi r^2 n_j^c(r) dr$.
%
By definition of $n_j^c$, 
$dn_j(r)$ provides the number 
of particles at distance $r$ which would be captured by the BH
if they did not collide with other particles in the surrounding
plasma. The account of collisions on particle propagation is
described by the equation $dn_j/dx = - \lambda_j n_\gamma$, where
$\lambda_j$ is the particle's mean free path in the medium. In the
case of collisions with photons in the thermal bath with temperature
$T$, the mean free path is given by
\be\label{lambda_j}
\lambda_j = \frac{1}{\sigma_{j\gamma} n_\gamma} \,
\sqrt{\frac{m_j}{T}} \, ,
\ee
where $\sigma_{j\gamma} = 8\pi\alpha^2/m_j^2$, $n_\gamma = 0.24 T^3$
is the number density of photons and the last factor accounts for
the necessity to make $\sqrt{m_j/T}$ collisions to transfer momentum
of the order of the thermal momentum $\sim \sqrt{m_j T}$.
The latter factor is absent for low momentum particles. Anyhow its
presence or absence does not significantly change the results. So, 
we multiply this quantity by $\exp(-r/\lambda_j)$ to take into account 
the effects of the scattering in the plasma. To estimate the capture 
rate we need to divide this quantity by the time of the propagation
from $r$ to zero. This time can be found by the integration 
of the equation of motion
$\ddot{r}_j =  - F_j/(2 r^2)$.
In principle, the result depends upon the initial velocity and direction.
However, an order of magnitude estimate for the time is
$t(r) = 2/3 \, r^{3/2} \, F_j^{-1/2}$.
The total number of captured particles ($p$ or $e$) per unit time 
is
\be\label{dot-N}
\dot{N}_j &=& 4 \pi C_j \int_0^\infty dr \frac{r^2}{t(r)} \exp(-r/\lambda_j) 
\int_0^{v_j^c(r)} dv v^2 \exp(-m_j v^2/2T) \nonumber\\
&=& 24 \sqrt{\pi} n_j F_j^2 
\left(\frac{m_j}{2T}\right)^{3/2} \int_0^\infty
\frac{dx}{x}\, \exp(-r_g x /\lambda_j) 
\int_0^1 du u^2 \exp \left( -\frac{m_j u^2 F_j}{2Tr_gx} \right) \, .
\ee
The main contribution to this integral comes from small $x$, 
where the exponent in the second term is equal to 1, so
\be
\dot{N}_j = 8 \sqrt{\pi} n_j F^2_j 
\left(\frac{m_j}{2T_j}\right)^{3/2} 
\ln \frac{\lambda_j}{r_g} \, .
\label{dot-N3}
\ee
The temperature of protons may be smaller than the temperature of 
electrons. The latter is usually equal to the temperature of the 
photon bath. The mean free paths $\lambda_j$ of $e$ and $p$ are 
also different, but their impact on the result is weak, because 
$\dot{N}_j$ depends on $\lambda_j$ logarithmically.

The equilibrium state would be reached if $\dot{N}_p = \dot{N}_e$. 
Keeping in mind that $F_p \approx 1 + 2q$ and 
$F_e \approx 1 + 2Kq$, we find that the equilibrium takes 
place for
\be
(1+ 2q)^2 (m_p/2T_p)^{3/2} \ln(\lambda_p/r_g)
\approx (1 + 2Kq)^2 (m_e/2T_e)^{3/2} \ln(\lambda_e/r_g) \, .
\label{q-eq1}
\ee
Let us assume $Kq \gg 1 \gg q$. In this case we get 
\be
4 K^2 q^2 \approx K^{3/2} 
\left(\frac{T_e}{T_p}\right)^{3/2} 
\frac{\ln(\lambda_p/r_g)}{\ln(\lambda_e/r_g)} \, .
\label{q-eq2}
\ee
Since $\lambda_p/\lambda_e \approx K^{5/2}$ 
(see eq.~(\ref{lambda_j})), at the temperature $T=1$~keV one finds
\be
\frac{\ln (\lambda_p/r_g)}{\ln(\lambda_e/r_g)} =
1 + \frac{\ln(\lambda_p/\lambda_e)}{\ln(\lambda_e/r_g)} 
\approx 1.7 \, .
\ee
Thus $K q \sim 100$ and $q \sim 0.1$ and the assumption 
$Kq \gg 1 \gg q$ is consistent. The accumulated electric 
charge at equilibrium is
\be
\alpha Q \approx \frac{r_g m_p}{2 K^{1/4}} 
\left(T_e/T_p\right)^{3/4} 
\sqrt{\frac{\ln(\lambda_p/r_g)}{\ln(\lambda_e/r_g)}}
\label{alpha-Q}
\ee
and, for $T_p=T_e$, we obtain $\alpha Q \approx 0.10 \, m_p r_g$.
The electric charge $Q$, and hence the efficiency of our
mechanism, strongly depends upon the temperatures of protons
and electrons. We expect that they are different for different
astrophysical environments. The case $T_p = T_e = T_\gamma$
is the most reasonable, because the coupling of electrons
and protons to the thermal bath of the surrounding photons is
strong enough to establish equality of their temperatures.
If it were not so, one may argue that $T_p$ is larger than $T_e$
because the accreting protons have virial temperature, but
electrons cool efficiently, so that $T_p/T_e \gg 1$ and $Q$
is much smaller than our estimate. That does not occur here,
because the accretion onto small BHs is quite different from the one 
onto the Solar-mass BHs. The Eddington
accretion for small BHs is typically several orders of magnitude
larger than the Bondi accretion, because the former is proportional 
to the BH mass, while the latter - to the square of the BH mass. For 
example, in the case of $10^{20}$~g BH inside a cloud with
density $10^{14}$~cm$^{-3}$, the Eddington to Bondi accretion 
rate is about $10^{-10}$. We should thus expect that there are
not many collisions between particles of the accreting matter
and the gravitational energy is not released efficiently, but
advected into the horizon where it is lost. The efficiency 
parameter $\eta$ is surely very low and $T_e$ is close to $T_p$.
Moreover, it is even more probable that there is no accretion 
disk at all: the gravitational radius of a $10^{20}$~g BH
is about $10^{-8}$~cm, i.e. the size of an atom. At such small
scales, quantum effects are important and they should prevent 
the formation of a ``regular'' disk. On the other hand, as already
said, most of the captured
particles are quickly swallowed by the BH because they emit
radiation to their orbit acceleration and thus lose angular 
momentum efficiently.

If the electric field at the horizon is strong enough, there 
is the possibility of particle pair production by vacuum breakdown
(or Klein instability)~\cite{pair-prod}. If the BH radius 
is much larger than the Compton wavelength of the electron
$1/m_e = 4 \cdot 10^{-11}$~cm, i.e. the BH mass is 
$M \gg 2 \cdot 10^{17}$~g, we can assume that the electrostatic 
field near horizon does not change with the distance
and the electron--positron production 
proceeds according to the Schwinger result~\cite{schwinger}~\footnote{If
this were not so, we would have to take into account 
a decrease of the field at the Compton wave length of electron. This
would lead to larger value of the critical field,
see the last paper in ref.~\cite{pair-prod}.}. The 
pair production probability per unit time and volume is
\be\label{eq-w}  
W = \frac{m_e^4}{\pi^2}\left(\frac{E}{E_c}\right)^2 
\sum_{n=1}^\infty \frac{1}{n^2} 
\exp \left( - \frac{n \sqrt{\pi} E_c}{2 E} \right) \, ,
\ee
where $E_c = m_e^2/e$. Even if $E \neq 0$ implies $W \neq 0$,
the production of particles in a uniform electrostatic field
is efficient only if $E$ is close to the critical value $E_c$.
In our case, the equilibrium electric field at horizon is
\be
E = \frac{\alpha Q}{r_g^2} \approx 0.10 \, \frac{m_p}{r_g}
\ee  
and $E_c/E = 0.9 \left(M/10^{20} \; {\rm g}\right)$.
The accretion rate of protons~(\ref{dot-N3}) is
\be
\dot N_p = 10^{16} \,\frac{n_p}{10^{10}/{\rm cm^3}}\,
\left(\frac{r_g}{10^{-8}{\rm cm}}\right)^2\,
\left( \frac{1\,\,{\rm keV}}{T_p}\right)^{3/2} 
\left[ \frac{\ln (\lambda_p/r_g)}{40}\right] \; {\rm s^{-1}}
\label{dot-Nfin}
\ee  
and since the value of the electric field at horizon is close
to $E_c$, the $e^+e^-$ pair production is fast 
and the rate is at the level of the proton accretion rate, that
is eq.~(\ref{dot-Nfin}). Then, the created positron 
would run away from the BH due to Coulomb repulsion, 
while electrons are predominantly
back--captured: the BH works actually as an efficient 
antimatter factory, converting protons into positrons. The 
energy of positrons at infinity is about 
$\alpha Q/r_g \approx 0.15 \, m_p \approx 140$~MeV.

{\sc Discussion --}
Small BHs with the mass at the level of $10^{20}$~g surely 
cannot be produced by the stellar collapse, 
but might be created in the early Universe, through many
possible mechanisms~\cite{pbhs, carr}. Then, if their mass is larger 
than $5 \cdot 10^{14}$~g, their lifetime exceeds the age of the 
Universe and they could still live somewhere today. 
The proposed mechanism of positron production works if the primordial
BH is surrounded by a ionized medium with the density of roughly 
$10^{10}$~particles/cm$^{3}$, which is about fourteen
orders of magnitude smaller than the density required in ref.~\cite{turolla}.
For example, the mechanism can operate in the neighborhood of 
the super--massive BH in the Galactic Center. Here one finds an 
ionized medium with temperature of roughly 1~keV and particle 
number density of the order of $10^8$~cm$^{-3}$ in a sphere of
radius $\sim 10^{16}$~cm around the central BH~\cite{melia}.
A more favorable environment is the hot and dense atmosphere 
around accreting compact stars: in the accretion disk, the 
temperature of the plasma can reach 10~keV and the particle 
number density can be as high as $10^{20}$~cm$^{-3}$~\cite{s-t}. 
Primordial BHs may be abundant around ``old compact objects'',
e.g. the final product of the first stars, because captured in 
the early Universe~\cite{spolyar}. They can still live today in 
the accretion disk of these objects because they cannot be swallowed 
quickly by their hosts like ordinary matter: they do not lose energy 
easily. The main mechanism of energy loss for small BHs in a plasma 
is via dynamical friction and the phenomenon is discussed in 
ref.~\cite{spolyar}. One can see that, for BHs with a mass around 
$10^{20}$~g and for reasonable values of density and temperature of 
the plasma, the time necessary to fall into the central object exceeds 
the Universe age.

In principle there are three observable phenomena induced
by the process of proton--to--positron transformation.

${\it i)}$ Observation of {\it MeV positrons in cosmic rays}: such positrons
would be hard to observe, because they would quickly stop and 
annihilate in the high density environment. Indeed the mean
free path of 100~MeV positrons due to scattering on the photon
bath with temperature $T$ is
\be\label{p-t-bath}
\lambda_e = \frac{1}{\sigma_{e\gamma}n_\gamma} 
\sqrt{\frac{E_e}{T}} \sim 10^5 \, 
\left(\frac{1 \; {\rm keV}}{T}\right)^3 \; {\rm cm} \, .
\ee
So they stop quite fast and annihilate with electrons.
Also, the cross section $e\gamma$ should be smaller than 
the Thompson one, but for 100~MeV positrons and 1~keV photons
the correction is of order one and does not change our 
conclusions.

${\it ii)}$ Observation of the {\it 0.511~MeV line from $e^+e^-$ annihilation}.
Here the problem is that the dense medium, where the small BHs
should be to efficiently transform protons into positrons, would
disperse this line to a continuos background. Still the line
might survive if the effective size of the dense region around 
the BHs is, at least in one direction, smaller than the photon 
mean free path. The latter can be estimated as
\be
\lambda_\gamma = \frac{1}{\sigma_{e\gamma} n_e} 
\sim 10^{14} \, \left(\frac{10^{10} 
\; {\rm cm^{-3}}}{n_e}\right) \; {\rm cm}
\ee
for 0.5~MeV photons due to scattering on the background electrons.
So, the possibility of observing the line depends crucially on $n_e$.
It is natural to wonder whether the mechanism can explain
the observed 511~keV line from the Galactic Center~\cite{511-exp},
where the expected annihilation rate is at the level of 
$10^{43}$~s$^{-1}$ and several different mechanisms have been
already proposed~\cite{511-prop}. From eq.~(\ref{dot-Nfin}), 
we see that the number of BHs should be
\be
N_{BH} \sim 10^{27} \, 
\left(\frac{10^{10} \; {\rm cm^{-3}}}{n_p}\right) \,
\left(\frac{10^{20} \; {\rm g}}{M}\right)^2 \,
\left(\frac{T_p}{1 \; {\rm keV}}\right)^{3/2} \, .
\ee
Since the total mass of non--baryonic matter in the Bulge cannot 
be larger than $10^{10}$~$M_\odot$~\cite{max-mass}, 
the small BHs have to live in regions
with $n_p > 10^{14}$~cm$^{-3}$. It is difficult to make
concrete predictions, as the number density of BHs and the
temperature and particle number density of the environment are 
unknown. The upper bound on $N_{BH}$ is given by the condition 
that their total mass cannot be larger than the maximum mass 
of dark matter (there are indeed no other constraints for BHs 
with masses in the range $10^{17} - 10^{27}$~g~\cite{carr}). 
Their distribution would be surely inhomogeneous 
and it is not unreasonable that they can be accumulated around 
compact bodies in the center of the Galaxy.

${\it iii)}$ A more realistic possibility is likely the observation of the
total gamma ray luminosity created by the annihilated 100~MeV
positrons. According to eq.~(\ref{dot-Nfin}), each BH emits
about $10^{16}$~erg/s for $n_p = 10^{14}$~cm$^{-3}$. If a total
amount of $10 - 100$~$M_\odot$ small BHs are accumulated into
a high density cloud/atmosphere, they may be observable as a
luminous object with the power of about $10^{30}-10^{31}$~erg/s.


{\sc Conclusion --} 
If primordial light black holes surrounded by a ionized
medium exist in the Galaxy, they could operate as an efficient
antimatter factories converting accreting protons into positrons. 
The charging of BH by the predominant influx of protons
in comparison to electrons is achieved because of the 
much larger proton mass and correspondingly much larger proton
mobility in the surrounding medium. If the black hole is small, 
the electrostatic field at the horizon can exceed the critical 
value of the vacuum stability and electron--positron pair production 
by Schwinger mechanism becomes efficient. Primordial black
holes with the mass in the range $10^{15} - 10^{20}$~g might live
today in the Universe, injecting positrons. If these black holes
live in an ionized high density medium, such as around the 
super--massive black hole in the Galactic Center or in the 
atmosphere of violent stars, the proton conversion
into energetic positrons would enrich cosmic rays by positrons
with energeis in the interval 1--100 MeV and gamma radiation of
similar in energies, which possibly makes the effect experimentally 
observable.


{\sc Acknowledgments --}
C.B. and A.A.P. are supported in part by NSF under grant PHY-0547794 
and by DOE under contract DE-FG02-96ER41005.


\end{document}